\documentclass[sigconf,screen]{acmart}

\usepackage{xspace}
\usepackage{color, colortbl}  

\usepackage{adjustbox}
\usepackage{siunitx}
\usepackage{tabularx} 
\usepackage{amsmath}  
\usepackage{multirow}
\usepackage{graphicx}    
\usepackage{subcaption}  
\usepackage{algorithm}
\usepackage{algorithmic}
\usepackage{soul}

\usepackage{xcolor}
\definecolor{dgrey}{gray}{0.2}

\definecolor{ao(english)}{rgb}{0.0, 0.5, 0.0}

\definecolor{violet}{HTML}{6a51a3}

\usepackage[english]{babel}  
\addto\extrasenglish{

}

\usepackage[capitalize]{cleveref} 
\usepackage[inline]{enumitem} 

\newcommand{\myparagraph}[1]{\paragraph*{\hspace*{-\parindent}\normalsize\bf#1}}
\AtBeginDocument{%
  }

\copyrightyear{2026}
\acmYear{2026}
\setcopyright{cc}
\setcctype{by}
\acmConference[CHIIR '26]{2026 ACM SIGIR Conference on Human Information Interaction and Retrieval}{March 22--26, 2026}{Seattle, WA, USA}
\acmBooktitle{2026 ACM SIGIR Conference on Human Information Interaction and Retrieval (CHIIR '26), March 22--26, 2026, Seattle, WA, USA}
\acmPrice{}
\acmDOI{10.1145/3786304.3788842}
\acmISBN{979-8-4007-2414-5/2026/03}

\begin{document}

\title{Characterizing Personality from Eye-Tracking: The Role of Gaze and Its Absence in Interactive Search Environments}

\author{Jiaman He}
\orcid{0009-0007-2817-7675}
\authornote{These authors contributed equally to this work, which was conducted at the National Institute of Informatics (NII)}
\affiliation{%
\institution{RMIT University}
\city{Naarm/Melbourne}
\country{Australia}
}
\email{jiaman.he@student.rmit.edu.au}

\author{Marta Micheli}
\orcid{0009-0003-4562-0334}
\authornotemark[1]
\affiliation{%
\institution{University of Turin}
\city{Torino}
\country{Italy}
}
\email{marta.micheli@unito.it}

\author{Damiano Spina}
\orcid{0000-0001-9913-433X}
\affiliation{%
\institution{RMIT University}
\city{Naarm/Melbourne}
\country{Australia}
}
\email{damiano.spina@rmit.edu.au}

\author{Dana McKay}
\orcid{0000-0001-7522-1842}
\affiliation{%
\institution{RMIT University}
\city{Naarm/Melbourne}
\country{Australia}
}
\email{dana.mckay@rmit.edu.au}

\author{Johanne R. Trippas}
\orcid{0000-0002-7801-0239}
\affiliation{%
\institution{RMIT University}
\city{Naarm/Melbourne}
\country{Australia}
}
\email{j.trippas@rmit.edu.au}

\author{Noriko Kando}
\orcid{0000-0002-2133-0215}
\affiliation{%
\institution{National Institute of Informatics}
\city{Tokyo}
\country{Japan}
}
\email{kando@nii.ac.jp}

\begin{abstract}

Personality traits influence how individuals engage, behave, and make decisions during the information-seeking process. However, few studies have linked personality to observable search behaviors. This study aims to characterize personality traits through a multimodal time-series model that integrates eye-tracking data and \textit{gaze missingness}--periods when the user's gaze is not captured. This approach is based on the idea that people often look away when they think, signaling disengagement or reflection. We conducted a user study with 25 participants, who used an interactive application on an iPad, allowing them to engage with digital artifacts from a museum. We rely on raw gaze data from an eye tracker, minimizing preprocessing so that behavioral patterns can be preserved without substantial data cleaning. From this perspective, we trained models to predict personality traits using gaze signals. Our results from a five-fold cross-validation study demonstrate strong predictive performance across all five dimensions: Neuroticism (Macro F1 = 77.69\%), Conscientiousness (74.52\%), Openness (77.52\%), Agreeableness (73.09\%), and Extraversion (76.69\%). The ablation study examines whether the absence of gaze information affects the model performance, demonstrating that incorporating missingness improves multimodal time-series modeling. The full model, which integrates both time-series signals and missingness information, achieves 10–15\% higher accuracy and macro F1 scores across all Big Five traits compared to the model without time-series signals and missingness. These findings provide evidence that personality can be inferred from search-related gaze behavior and demonstrate the value of incorporating missing gaze data into time-series multimodal modeling.

\end{abstract}



\maketitle

\section{Introduction}
\label{sec:intro}

Consider the different ways people behave when searching for information. Some are easily distracted, while others display a strong sense of curiosity. Certain individuals prefer to skim ahead to the end, whereas others move back and forth through the material to piece together a deeper understanding. The motivations for searching also vary: some seek confirmation of what they already believe~\cite{boonprakong2025hci}, while others are driven by the desire to learn something new~\cite{kang2009wick}. Exploring these different behaviors is essential for advancing areas such as recommendation systems~\cite{hu2011enhancing, onori2016comparative, yusefi2018improving, dhelim2022survey}, personalization~\cite{ho2008personalization, javadi2026CHARISMA}, understanding confirmation bias~\cite{melinder2020personality}, or LLM-based simulations~\cite{10.1145/3726302.3730238, he2025can, leng2025agentsense, zerhoudi2026simulation}.
Among the many factors that shape these behaviors, personality traits stand out as a foundational influence, impacting human patterns of thought, preference, and decision-making across contexts~\cite{karumur2018personality}.

Personality traits fundamentally shape how and why people seek information. Curious and open individuals often seek intellectual engagement, while those high in neuroticism or intolerance of uncertainty tend to seek reassurance in negative contexts~\cite{jach2022people}.
The Big Five model~\cite{mccrae1987validation}, which includes Openness, Conscientiousness, Extraversion, Agreeableness, and Neuroticism, is widely used in personalization research. Studies link personality to information seeking~\cite{heinstrom2003five} and web search behaviors~\cite{ashkanasy2007effects}, showing its potential for personalized information retrieval (IR). However, while the importance of psychological dimensions in interaction and search has long been acknowledged as a major challenge in IR~\cite{belkin1990cognitive}, the impact of the systematic exploration personality on IR behaviors and its data-driven modeling remains in its early stages.

A direction for modeling personality in IR lies in eye-tracking data. 
With mobile and wearable devices (e.g., Tobii Pro Glasses 3~\cite{tobiiProGlasses3}, Apple Vision Pro~\cite{appleVisionPro}) now featuring eye-tracking capabilities~\cite{krafka2016eye}, it is possible to collect such data in real-world settings. Eye movements provide insights into cognitive and attentional processes even before a user makes a decision~\cite{he2025characterising}. Previous work has investigated links between personality and gaze patterns~\cite{hoppe2018eye,berkovsky2019detecting}, though studies focusing specifically on personality inference from eye movements during digital search interactions remain limited~\cite{chen2023eye,millecamp2021classifeye,woods2022twenty}. In addition, eye-tracking data also present challenges; for instance, \textit{data missingness} occurs frequently when users are not looking at the screen, and noise often requires complex preprocessing. We address these issues by interpreting missingness--the absence of gaze--as informative signals and by using raw gaze data obtained directly from the eye tracker (Section~\ref{apparatus}), avoiding additional cleaning steps in order to capture behavioral patterns without extensive preprocessing.

Building on this perspective, we investigate the role of viewing behaviors in recognizing users' Big Five personality traits during search tasks. Our user study uses a GUI-based application displaying museum items, enabling the collection of rich data on interactive search behaviors, including gaze patterns and engagement dynamics.
\noindent Our main contributions are as follows:

\begin{enumerate}[topsep=0pt]

    \item We propose a multimodal time-series model that incorporates missing eye-tracking data as behavioral cues, integrating gaze and pupil signals to capture both attention patterns and underlying cognitive states.

    \item We show that users’ Big Five personality traits can be predicted from eye-tracking data in complex search environments, achieving a classification performance across all five dimensions: Neuroticism (Macro F1 = 77.69\%), Conscientiousness (74.52\%), Openness (77.52\%), Agreeableness (73.09\%) (see Table~\ref{tab:bigfive_results}).

    \item We present findings that characterize how users search and explore digitized museum collections, based on a controlled user study ($N=25$) involving 
    a graphical digital interface. 
   %
\end{enumerate}
\section{Related Work}
\label{sec:relatedwork}
\subsection{Personality Traits and Search Behavior}

Personality traits describe stable patterns in how individuals think, feel, and behave~\cite{john1999big}. These traits shape emotional and motivational processes, influencing how people seek, process, and use information~\cite{sanderson2007examining}. Prior research indicates that Big Five personality traits are associated with differences in information search
behaviors~\cite{jach2022people}. 

Curiosity reflects a drive to explore novelty, complexity, or ambiguity~\cite{kashdan2018five}. It can be expressed as deprivation sensitivity, where individuals are motivated to close knowledge gaps, or as joyous exploration, where the act of learning itself is intrinsically rewarding~\cite{kashdan2018five}. By contrast, intolerance of uncertainty describes the tendency to experience uncertain situations as threatening, distressing, or undesirable~\cite{carleton2007fearing}.

Curiosity—particularly joyous exploration—is usually understood as part of openness/intellect, which captures differences in creativity, imagination, and intellectual engagement~\cite{silvia2020looking,kashdan2018five}. Intolerance of uncertainty aligns more closely with neuroticism, reflecting vulnerability to negative affect and worry~\cite{jach2019fear,belkin1990cognitive}. Together with extraversion, agreeableness, and conscientiousness, these dimensions form the Big Five model of personality~\cite{markon2009hierarchies}. 

These traits play an important role in information-seeking behavior. Individuals high in joyous exploration may search broadly and openly, driven by intrinsic interest, while those high in deprivation sensitivity may pursue information more urgently to resolve perceived knowledge gaps. Conversely, individuals with high intolerance of uncertainty may engage in searching defensively, focusing on information that reduces ambiguity or confirms certainty, while avoiding open-ended exploration. In this way, personality traits can be seen as probabilistic predictors of how people approach information environments: shaping whether they explore widely, focus narrowly, avoid uncertainty, or embrace it.

From a dynamic perspective, personality traits should not be viewed as fixed responses but as distributions of tendencies across time and context~\cite{fleeson2009implications}. For example, someone high in intolerance of uncertainty will not resist ambiguity in every instance but will do so more frequently and with greater intensity than someone lower on that trait. Similarly, individuals high in curiosity will tend to experience more frequent and stronger motivation to seek new information~\cite{jach2022people}. Thus, understanding personality traits provides valuable insight into the variability of information-seeking strategies across individuals.

In our study, we incorporate the Big Five personality traits~\cite{mccrae1987validation}, commonly summarized by the acronym \textbf{OCEAN}: Openness (O), Conscientiousness (C), Extraversion (E), Agreeableness (A), and Neuroticism (N). We use eye-tracking data to examine how these traits relate to user behavior.  


\vspace{-0.1in}
\subsection{Eye Tracking and Personality Traits}

Eye-tracking has been widely used to investigate people’s cognitive states and their relationship to information-seeking behavior~\cite{eickhoff2015eye,cole2013inferring,he2025characterising}. Research suggests that individuals with different personality traits may follow distinct eye movement patterns when searching for information~\cite{al2017impact}. More broadly, prior work has established that personality traits can influence gaze control and visual attention~\cite{matsumoto2010factors, rauthmann2012eyes,al2017impact,sarsam2023influence}. This has motivated a series of studies exploring whether personality can be inferred directly from eye-tracking data.

\citet{hoppe2018eye} examined the Big Five traits during everyday activities. Participants wore head-mounted eye trackers while walking and shopping, and the resulting gaze data were used to train random forest classifiers. Their models achieved above-chance prediction of Extraversion, Neuroticism, Agreeableness, and Conscientiousness. \citet{berkovsky2019detecting} extended this work to controlled settings, proposing a framework for predicting psychological characteristics—including the Big Five—based on passive viewing of images and videos. Using data from 21 participants, they applied supervised machine learning techniques and achieved encouraging levels of accuracy.

Other studies focused on interactive systems. \citet{chen2023eye} collected gaze data during product selection tasks with recommender systems, applying multiple classifiers and feature selection methods to predict personality. \citet{millecamp2021classifeye} studied gaze patterns in a music recommender system, with models predicting traits such as Openness to Experience, though their accuracy was not yet sufficient for practical applications.

Expanding to immersive contexts, \citet{khatri2022recognizing} examined user behavior in a virtual reality shop, incorporating three-dimensional gaze features to predict the Big Five. \citet{woods2022twenty} investigated social media use, tracking participants as they browsed their Facebook News Feeds. Using only 20 seconds of gaze data per user, classification produced acceptable results for Extraversion and Conscientiousness, though performance was weaker for other traits.

Collectively, these studies highlight the dual role of eye tracking: it not only provides insights into cognitive and information-seeking processes but also offers a potential signal for inferring personality. Personality-related differences in gaze behavior suggest that personality-aware models may help explain why individuals vary in how they explore, attend to, and process information.

A practical challenge is that eye-tracking data often contain noise, such as blinks or missing samples when users look away~\cite{ji2024characterizing}. Previous studies have typically handled these gaps by interpolation, replacement, or removal~\cite{winn2018best,franzen2022individual,mathot2018safe,blumenfeld2002neuroanatomy,he2025characterising}. However, we argue that instances of looking away are themselves informative, as they may reveal valuable aspects of human behavior. To leverage this, we propose a novel method that preserves these missing segments. Our approach, detailed in~\autoref{Missing Data Aware Network for Personality Prediction}, integrates the missing data into a time-series framework to better capture behavioral patterns.

\vspace{-0.1in}
\subsection{Time-Series Modeling and Missing Data}

\citet{cole2015user} analyzed time-series activity patterns to study search behavior. Similarly, eye-tracking data can be represented as a multivariate time series, capturing variables such as gaze coordinates, pupil size, and gaze velocity continuously over time \cite{langkvist2014review}. At each timestep, a vector of measurements is recorded, and the sequence of these vectors traces the dynamics of visual attention as it evolves moment by moment.

Most prior work on personality inference from gaze has reduced these streams to aggregate features, such as fixation counts, blink rates, or average saccade lengths\cite{berkovsky2019detecting, hoppe2018eye, millecamp2021classifeye,woods2022twenty}. While useful, such handcrafted summaries discard the fine-grained temporal patterns, such as prolonged fixations, systematic scanning, or rhythmic shifts in attention, that may carry diagnostic information about personality. By contrast, sequence models such as recurrent neural networks \cite{hochreiter1997long}, convolutional temporal encoders \cite{wang2017time}, and more recently Transformers \cite{lim2021temporal} are designed to capture dependencies across time, making them well suited for modeling how gaze evolves during complex tasks.

To motivate this shift, it is helpful to look at parallel domains where similar challenges arise. In human activity recognition (HAR), wearable sensors generate continuous multivariate streams (e.g., accelerometer, gyroscope) that are best understood as sequences \cite{ordonez2016deep}. Likewise, in healthcare, physiological signals such as ECG or PPG are modeled as multivariate time series to capture patterns in heart rhythms or respiration \cite{hong2020opportunities}. In both cases, moving from handcrafted features to sequence modeling has led to significant improvements in predictive accuracy and robustness \cite{holmqvist2011eye}. These domains demonstrate the value of treating behavioral signals as structured temporal data—an approach we adopt for eye-tracking in information retrieval.

Another challenge with eye-tracking is the prevalence of missing data. Blinks, momentary loss of calibration, or glances away from the screen introduce gaps in the signal. Earlier studies often discarded these segments or filled them in with simple interpolation \cite{hoppe2018eye,millecamp2021classifeye}. However, research in healthcare time series has shown that missingness can itself be informative \cite{lipton2016modeling}: the duration and frequency of gaps may reflect underlying behavioral or physiological states. Modern approaches augment sequential inputs with masking vectors and temporal gap features, allowing models to distinguish between transient noise or short-term dropouts and more prolonged periods of missingness in the signal \cite{che2018recurrent}.

Despite the success of these methods in other domains, they have not been applied to eye-tracking for personality prediction. Our study builds on these insights by proposing a missing-data-aware sequence modeling framework, which treats gaze not only as a multivariate temporal signal but also as a behavioral record where the absence of data may carry meaningful information about individual differences in attention and personality.
\section{Methodology}
\label{sec:Experiment}
\subsection{Experimental Setup}

\begin{figure}[htbp]
    \centering
    \begin{subfigure}[b]{0.45\textwidth}
        \centering
        \includegraphics[width=\textwidth]{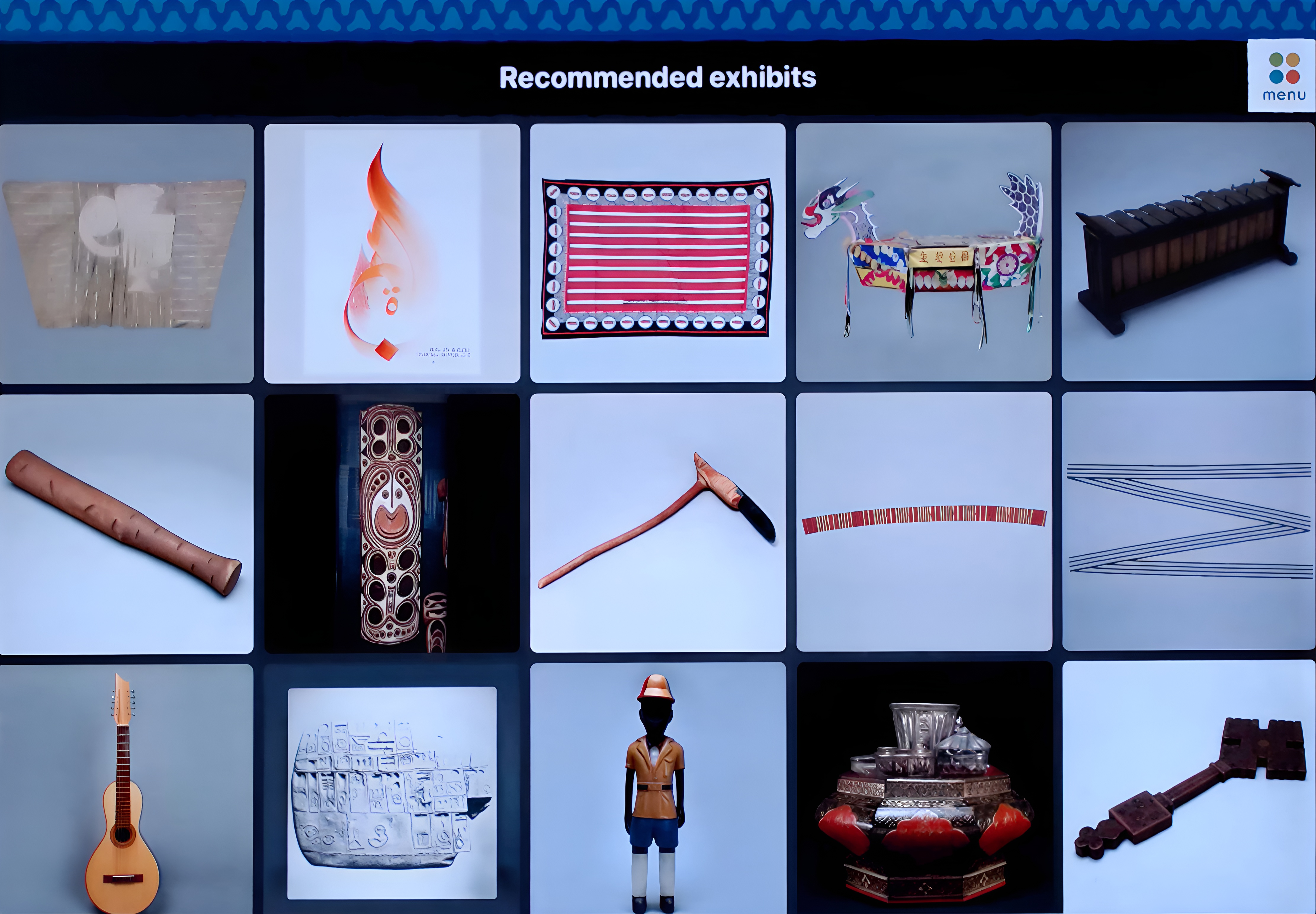}
        \caption{Example of image grid page}
        \label{fig:photo grids}
    \end{subfigure}
    \hfill
    \begin{subfigure}[b]{0.45\textwidth}
        \centering
        \includegraphics[width=\textwidth]{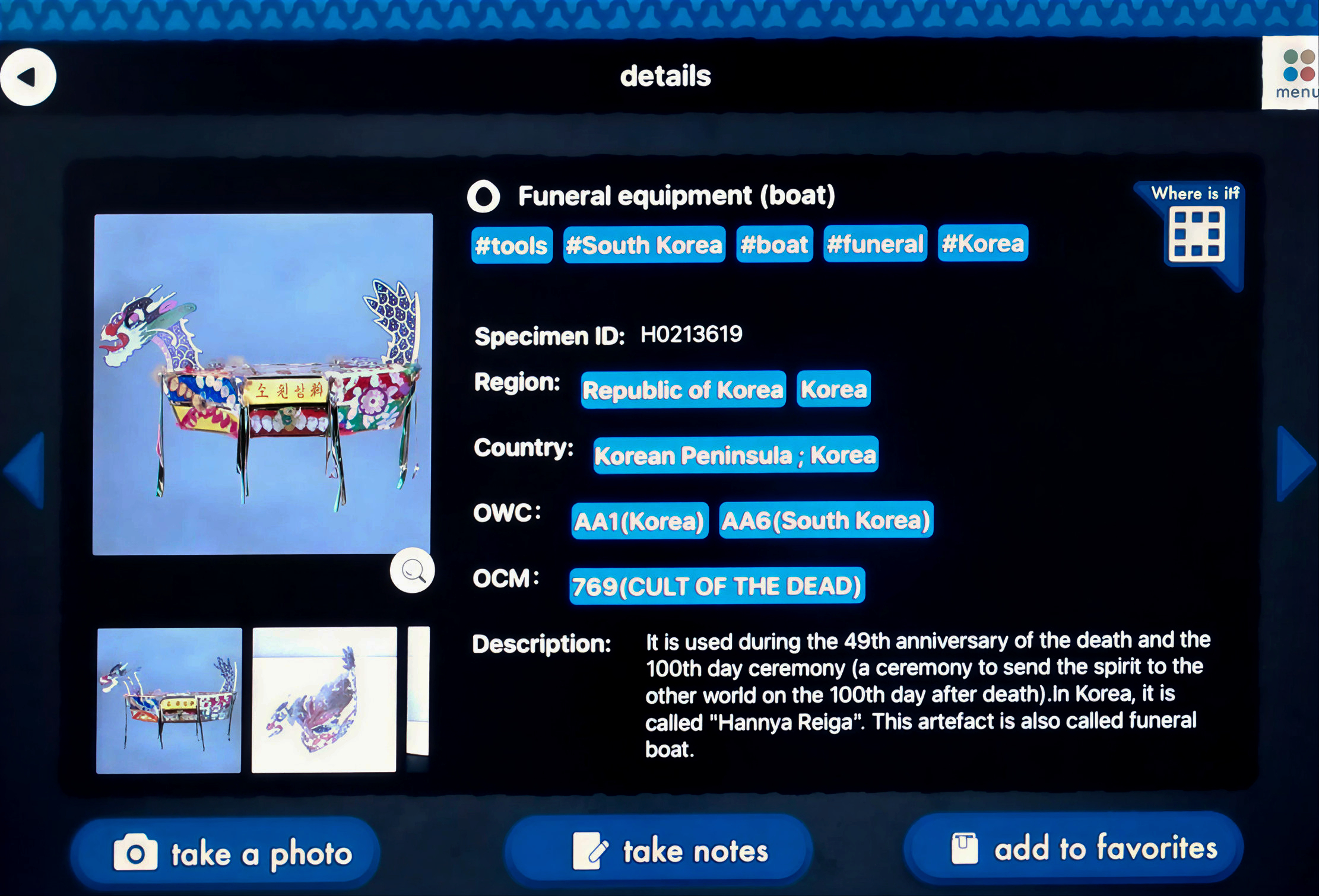}
        \caption{Example of description page}
        \label{fig:description page}
    \end{subfigure}
    
    \caption{Overview of the application interfaces}
    \label{fig:Overview of the application interfaces}
\end{figure}

Experiments were conducted using the \textit{Minpaku Guide}~\cite{shoji2021museum}, an iPad application developed for the National Museum of Ethnology in Osaka, Japan. For this study, the English version of the app was employed. The application provides extensive content related to the museum’s artifacts and is based on an \textit{ostensive search model~\cite{campbell1996ostensive}}. The interface follows the same structural and functional principles as a search engine results page (SERP): users begin by browsing a grid of artifact photos, which they can select to view detailed information and scroll to explore related items~\cite{shoji2021museum}. The app consists of multiple page types, including (i) modular image grid (shown in~\autoref{fig:photo grids}), (ii) individual page, which contains description about a single museum object (shown in~\autoref{fig:description page}), and (iii) map-based views.

\subsubsection{Procedure}


The study followed a structured procedure: participants gave informed consent, completed a pre-task questionnaire and short training, performed the search task, and concluded with a post-questionnaire and brief interview with stimulus recall.

This paper uses the data collected in the following two phases:

\begin{enumerate}
    \item \textbf{Personality Assessment.} As part of pre-task questionnaire, participants completed the \textit{BFI-44 questionnaire}~\cite{donahue1991big} in English to measure their Big Five personality traits. This self-reporting instrument has been shown to be reliable~\cite{arterberry2014application}.  
    \item \textbf{Search Task.} Participants explored the Minpaku Guide and selected five items they liked most, marking them as \textit{favorites}. No time limit was imposed to avoid biasing exploratory behavior. In addition to user's eye movement (see Section~\ref{apparatus}), the application also logged detailed interaction data, including item offsets and vertical scroll positions, since all the pages were scrollable.
\end{enumerate}

This study was reviewed and approved by the Institutional Review Board (IRB) of National Institute of Informatics (NII), Japan. 


In preparing the application for the task, we ensured that the quality and arrangement of the images on the first page were appropriate for the experiment. This page, which participants encountered first when using the Minpaku Guide, displayed multiple artifacts in a grid layout. To maximize heterogeneity, we selected images from a larger set using an agglomerative clustering method ($K = 50$) applied to image features. The features were extracted with ResNet50~\cite{he2016deep}, a deep learning convolutional neural network widely used in image processing research. The items displayed in the Minpaku Guide were identical for all participants.



\subsubsection{Participants}
\label{participants}

Twenty-five participants took part in the experiment, representing different nationalities and ranging in age from 18 to 34 years. Participants were recruited using a snowball sampling strategy.
Participants were informed about the purpose and procedures of the study, including its expected duration, potential risks, and policies regarding data storage and usage to safeguard privacy. Participants were also explicitly told that they could withdraw from the study at any point without penalty. All participants provided written informed consent prior to the experiment.

\subsubsection{Apparatus}
\label{apparatus}

Eye tracking was conducted using a Tobii Pro Nano at a sampling rate of 60 Hz. The device, connected to a Dell Precision 7550 computer, was positioned above an iPad Pro 2022 to avoid obstructing participants during interaction. Each participant sat approximately 50 cm from the iPad.

Because a video capture card was unavailable, the screen activity of the iPad application, along with participants’ gaze trajectories, was recorded using a Logitech C920 webcam. The webcam feed was shown on a separate monitor that was not visible to the participant.

Tracking was performed for both eyes and their recorded data was used for analysis. Recordings had a resolution of 1024 × 576 pixels with an average accuracy of 0.52°. Ambient lighting was kept consistent across participants to avoid effects on pupil dilation.


\subsection{Missing-Data-Aware Network for Personality Prediction}
\label{Missing Data Aware Network for Personality Prediction}

We propose a missing-data-aware framework for predicting personality traits from eye-tracking signals. First, raw gaze coordinates, pupil size, and gaze velocity are represented as multivariate time series segmented into fixed-length windows (Section \ref{subsec: eye tracking multivariate}). Second, missingness is modeled with binary masks and temporal gap features, treating absent data as informative signals rather than noise (Section \ref{subsec: sequence modeling}). Finally, the augmented sequences are processed by a bidirectional LSTM, which captures temporal dependencies and produces window-level predictions (Section \ref{subsec: sequence modeling}). Together, these components form a pipeline that leverages both eye-movement patterns and missing data for personality prediction. An overview of the full pipeline is presented in Algorithm~\ref{alg:missing_data_network}.

\begin{algorithm}[t]
\caption{Training Missing-Data-Aware Network for Personality Prediction}
\label{alg:missing_data_network}
\begin{algorithmic}[1]
\REQUIRE Raw eye-tracking sequence $\{(g_t^x, g_t^y, p_t)\}_{t=1}^T$, sampling period $\Delta t$, window length $L$, number of epochs $E$
\ENSURE Trained BiLSTM parameters $\theta = \{W, b\}$

\STATE Compute features:
  \begin{itemize}
    \item Compute gaze velocity $v_t = \|\mathbf{g}_t - \mathbf{g}_{t-1}\|_2 / \Delta t$ ($v_1=0$).
    \item Normalize gaze coordinates: $\bar g_t^x, \bar g_t^y \in [-1,1]$.
    \item Standardize pupil diameter $p_t$ and velocity $v_t$ with z-scores.
  \end{itemize}

\STATE Handle missingness:
  \begin{itemize}
    \item Replace NaN with $0$.
    \item Construct binary mask $\mathbf{m}_t \in \{0,1\}^d$.
    \item Update temporal gaps $\boldsymbol{\Delta t}_t$ recursively.
  \end{itemize}

\STATE Build augmented input $\mathbf{f}_t = [\tilde{\mathbf{x}}_t, \mathbf{m}_t, \boldsymbol{\Delta t}_t] \in \mathbb{R}^k$.
\STATE Partition sequence into overlapping windows $\mathbf{W}^{(n)}_k \in \mathbb{R}^{L \times k}$.

\FOR{epoch $=1$ to $E$}
  \FOR{each window $\mathbf{W}^{(n)}_k$ with label $y^{(n)}$}
    \STATE Encode window with BiLSTM $\rightarrow \{\mathbf{h}_t\}_{t=1}^L$.
    \STATE Summarize: $\mathbf{z} = [\overrightarrow{\mathbf{h}}_L; \overleftarrow{\mathbf{h}}_1]$.
    \STATE Predict with softmax: $\hat{y} = \text{softmax}(W\mathbf{z}+b)$.
    \STATE Compute loss: $\mathcal{L} = -\sum_{c=1}^C \mathbf{1}[y=c] \log \hat{y}_c$.
    \STATE Update parameters $\theta$ with gradient descent.
  \ENDFOR
\ENDFOR
\end{algorithmic}
\end{algorithm}

\subsubsection{Eye-Tracking Data as a Multivariate Time Series}
\label{subsec: eye tracking multivariate}
Let $f_s$ denote the sampling frequency (Hz), and $\Delta t = 1/f_s$ the sampling period. 
That is, the eye tracker records $f_s$ samples per second, with $\Delta t$ seconds between two consecutive recordings.  

For a recording session of length $T$ samples, we represent the eye-tracking stream as a multivariate time series, where each sample is represented as follows:
\[
\mathbf{x}_t = \big[g_t^x,\; g_t^y,\; p_t,\; v_t\big]^\top \in \mathbb{R}^d, 
\quad t=1,\dots,T,\qquad d=4,
\]
where $g_t^x,g_t^y$ are the horizontal and vertical gaze coordinates on the display, 
$p_t$ is pupil diameter, and $v_t$ is gaze velocity. 
Each $\mathbf{x}_t$ is therefore a four-dimensional snapshot of eye behavior at time $t$. 

The gaze velocity is computed from the gaze position vector $\mathbf{g}_t = (g_t^x,g_t^y)^\top$ as 
\[
v_t \;=\; \frac{\|\mathbf{g}_t - \mathbf{g}_{t-1}\|_2}{\Delta t}, \quad t \ge 2,
\]
with $v_1=0$ by convention. This captures how far the eyes moved between consecutive samples, divided by the elapsed time.  

\paragraph{Session and window representation.}
We collect the full session as 
\[
\mathbf{X}=\big[\mathbf{x}_1,\mathbf{x}_2,\dots,\mathbf{x}_T\big] \in \mathbb{R}^{d\times T},
\]
with timestamps $\tau_t = (t-1)\Delta t$. 
Here, $\mathbf{X}$ is arranged so that each column corresponds to one moment in time ($t=1,\dots,T$), 
and each row corresponds to one of the $d=4$ recorded features 
(horizontal gaze $g^x$, vertical gaze $g^y$, pupil size $p$, velocity $v$). 
Thus, $\mathbf{X}$ can be seen as a compact table of the entire session: 
moving across columns follows the sequence of time steps, while moving down rows inspects the different measurements collected at that time.

For model training, this long sequence is divided into overlapping subsequences (windows) of fixed length $L$:
\[
\mathbf{W}_k = \big[\mathbf{x}_{s_k}, \mathbf{x}_{s_k+1}, \dots, \mathbf{x}_{s_k+L-1}\big] \in \mathbb{R}^{d \times L}, 
\quad k=1,\dots,K,
\]
where $s_k$ is the start index of the $k$-th window and $K$ is the total number of extracted windows. 
Each $\mathbf{W}_k$ is therefore a short clip of the session that spans $L$ consecutive timesteps 
but keeps all $d$ features at each step. 
In practice, this sliding-window procedure allows the model to learn from many local fragments of behavior, 
capturing recurring gaze patterns that may be predictive of personality traits.

\paragraph{Normalization.}
Let $(W,H)$ denote the screen width and height in pixels, and $(g_t^x,g_t^y)$ the raw gaze position 
recorded at time $t$ in pixel coordinates. 
To remove dependence on the specific display size, gaze coordinates are rescaled to a zero-centered, unit-square system: $\bar g_t^x \;=\; \frac{2g_t^x}{W} - 1,\qquad 
\bar g_t^y \;=\; \frac{2g_t^y}{H} - 1,$
so that $(\bar g_t^x,\bar g_t^y) \in [-1,1]^2$. 
After this transformation, the center of the screen corresponds to $(0,0)$, 
the left and right edges correspond to $-1$ and $+1$ on the $x$-axis, 
and the top and bottom edges correspond to $+1$ and $-1$ on the $y$-axis. 
This mapping ensures that gaze positions are expressed in a common reference frame across devices and participants.  

For the pupil diameter, let $p_t$ be the raw measurement at time $t$, 
$\mu_p$ the mean pupil diameter across the training set, 
and $\sigma_p$ the corresponding standard deviation. 
We standardize pupil diameter using z-score normalization: $\tilde p_t \;=\; \frac{p_t - \mu_p}{\sigma_p}.$

Similarly, let $v_t$ denote the raw gaze velocity, 
with $\mu_v$ and $\sigma_v$ the mean and standard deviation of velocity over the training set. 
Velocity is standardized in the same way: $\tilde v_t \;=\; \frac{v_t - \mu_v}{\sigma_v}.$

This standardization rescales $p_t$ and $v_t$ to have zero mean and unit variance, 
so that the model focuses on relative fluctuations (e.g., dilations above or below a typical pupil size, 
or faster versus slower gaze shifts) rather than absolute raw values, 
which may vary considerably across individuals.

\paragraph{Final feature vector.}
Unless otherwise noted, the input at each timestep $t$ is the normalized feature vector
\[
\tilde{\mathbf{x}}_t \;=\; \big[\bar g_t^x,\; \bar g_t^y,\; \tilde p_t,\; \tilde v_t\big]^\top \in \mathbb{R}^4,
\]
which stacks together the four processed measurements: normalized horizontal gaze position $\bar g_t^x$, 
normalized vertical gaze position $\bar g_t^y$, standardized pupil diameter $\tilde p_t$, 
and standardized gaze velocity $\tilde v_t$.  


\subsubsection{Modeling Missingness in Eye-Tracking Data}
\label{subsec: missing data modeling}
Eye-tracking data collected under naturalistic conditions often contain substantial missing values. 
This missingness arises when the participant blinks, looks away from the screen, or when the tracker momentarily loses calibration. 
In the raw data, such missing entries are recorded as \texttt{NaN}. 
Since the model cannot operate directly on \texttt{NaN} values, we replace them with zeros before training. 
However, the value $0$ can also be a valid observation (for example, a gaze coordinate of $0$ corresponds to the center of the screen). 
To prevent confusion between true zeros and placeholders for missing values, we introduce an explicit binary validity mask.

\paragraph{Binary validity masks.}
For each feature dimension $j \in \{1,\dots,d\}$ at time $t$, 
we define a binary indicator variable
\[
m_t^j =
\begin{cases}
1, & \text{if feature $j$ has a valid observed value at time $t$}, \\
0, & \text{if feature $j$ is missing (recorded as \texttt{NaN} in the raw data)},
\end{cases}
\]
where $d=4$ in our case (horizontal gaze, vertical gaze, pupil size, and velocity).  
These indicators are then collected into a mask vector
\[
\mathbf{m}_t = [m_t^1, m_t^2, \dots, m_t^d]^\top \in \{0,1\}^d.
\]

The role of this mask is to preserve the distinction between two different situations:  
(1) a feature truly takes the value $0$ (e.g., $\bar g_t^x = 0$ meaning the gaze is exactly at the horizontal center of the screen),  
and (2) the original data at that position was missing and has been replaced with $0$ only as a placeholder so the model can process the input.  
Without the mask, these two cases would be indistinguishable.  

Concretely, if the horizontal gaze coordinate is missing at time $t$, we set its value in $\tilde{\mathbf{x}}_t$ to $0$ but also set $m_t^1 = 0$.  
If the gaze is genuinely at the screen center, then the value is also $0$ but the mask records $m_t^1 = 1$.  
Thus, the pair $(\tilde{\mathbf{x}}_t, \mathbf{m}_t)$ allows the model to differentiate between a true zero measurement and a zero that only indicates missingness.

\paragraph{Temporal gap encoding.}
In addition to the binary masks, we record how long each feature has been continuously missing. 
For feature $j$ at time $t$, we define a temporal gap variable $\Delta t_t^j$ that is updated recursively as
\[
\Delta t_t^j =
\begin{cases}
0, & \text{if $m_t^j=1$ (feature $j$ is observed at time $t$)}, \\
\Delta t_{t-1}^j + \Delta t, & \text{if $m_t^j=0$ (feature $j$ is missing at time $t$)},
\end{cases}
\]
where $\Delta t$ is the sampling period.  

Thus, $\Delta t_t^j$ counts how long feature $j$ has been missing up to time $t$.  
Whenever a valid measurement is observed ($m_t^j=1$), the gap resets to $0$.  
When the feature remains missing across consecutive timesteps ($m_t^j=0$), the gap grows by $\Delta t$ each time step.  

This variable provides temporal context for the missingness:  
\begin{itemize}
  \item If $\Delta t_t^j$ is small (close to $0$), the absence is likely due to a short interruption such as a blink or a brief calibration error.  
  \item If $\Delta t_t^j$ grows large, it indicates a prolonged disengagement, for example the participant looking away from the screen for several seconds.  
\end{itemize}

By including $\Delta t_t^j$ as an explicit feature, the model can distinguish between short, transient dropouts and longer episodes of missing data, which may carry different behavioral meanings.

\paragraph{Extended feature representation.}
At each timestep $t$, we construct an augmented input vector

\[
\mathbf{f}_t = \big[\tilde{\mathbf{x}}_t,\; \mathbf{m}_t,\; \boldsymbol{\Delta t}_t\big],
\]
where:
\begin{itemize}
  \item $\tilde{\mathbf{x}}_t \in \mathbb{R}^d$ are the normalized feature values 
  (horizontal and vertical gaze coordinates, standardized pupil size, and standardized velocity),
  \item $\mathbf{m}_t \in \{0,1\}^d$ is the binary validity mask, indicating for each feature whether the value at time $t$ was truly observed ($1$) or was originally missing and replaced by a placeholder ($0$),
  \item $\boldsymbol{\Delta t}_t = [\Delta t_t^1, \dots, \Delta t_t^d]^\top \in \mathbb{R}^d$ contains the temporal gap variables, which record how long each feature has been continuously missing.
\end{itemize}

This yields an augmented representation of dimension $k = 3d$: for each of the $d$ base features, we include its normalized value, a validity flag, and a temporal gap duration.  

The motivation for this augmentation is threefold:  
\begin{enumerate}
  \item The model can \emph{ignore invalid entries} by using the mask $\mathbf{m}_t$ while still retaining information about which features were missing.  
  \item The temporal gap variables $\boldsymbol{\Delta t}_t$ allow the model to capture behavioral patterns such as the difference between a short blink (a brief gap) and sustained disengagement (a long gap).  
  \item By combining observed values with missingness structure, the model can potentially learn \emph{personality-related regularities}, for example, that certain individuals tend to look away more often or for longer periods, which may be predictive of traits like neuroticism or conscientiousness.
\end{enumerate}

\subsubsection{Sequential Modeling for Personality Prediction}
\label{subsec: sequence modeling}
Personality prediction from eye-tracking is framed as a sequence-to-label task: 
the model receives a short sequence of eye-tracking data and must predict the participant’s personality traits.  

Given an augmented feature window
\[
\mathbf{W}^{(n)}_k = [\mathbf{f}_{s_k}, \mathbf{f}_{s_k+1}, \dots, \mathbf{f}_{s_k+L-1}]^\top 
\in \mathbb{R}^{L \times k},
\]
we stack $L$ consecutive augmented vectors $\mathbf{f}_t \in \mathbb{R}^k$ (normalized features, validity masks, and temporal gaps).  
Here, $n$ indexes the participant, $s_k$ is the start index of the $k$-th window, and $k$ is the feature dimension.  

The prediction target is $y^{(n)} \in \{1,2,3\}^5,$
a 5-dimensional vector for the Big Five traits (Openness, Conscientiousness, Extraversion, Agreeableness, Neuroticism), 
where each entry takes values $1,2,3$ for \emph{Low}, \emph{Medium}, and \emph{High}. 
For the classification setup, participants’ continuous personality scores (originally measured on a 1–5 scale) were 
partitioned into three groups using a quantile-based procedure following \cite{Saboundji2024personality}. For each trait, cut points were set at the 33rd and 
66th percentiles of its empirical distribution, so that each group contained roughly the same number of participants. 
This procedure yielded balanced class sizes and provided clear separation among low, medium, and high scorers. All windows from participant $n$ share the same label $y^{(n)}$, so the task 
is to learn a mapping from temporal sequences $\mathbf{W}^{(n)}_k$ to the trait categories.

\paragraph{Sequential representation learning.}
Prior work often summarized windows into handcrafted statistics, 
such as average fixation duration, blink counts, or velocity variance.  
While useful, such summary features discard the temporal ordering of the signal.  
Here we instead retain the sequence itself, enabling the model to exploit fine-grained temporal dependencies—such as sustained fixations, 
repeated scanning motions, or short pupil dilations—that may be characteristic of personality traits.  

\paragraph{BiLSTM encoder.}
To capture temporal dependencies in the data, we employ a bidirectional Long Short-Term Memory (BiLSTM) network. 
An LSTM is a type of recurrent neural network that is designed to process sequences one element at a time, 
while retaining information from previous steps through a hidden state. 
This makes it well suited to modeling time-series signals such as eye movements, 
where the current behavior depends strongly on what came before.  

At each timestep $t$, the hidden state is updated as
\[
\mathbf{h}_t = \text{BiLSTM}(\mathbf{f}_t, \mathbf{h}_{t-1}), 
\quad \mathbf{h}_t \in \mathbb{R}^h,
\]
where $\mathbf{f}_t$ is the augmented feature vector at time $t$ and $h$ is the dimensionality of the hidden state.  

Unlike a standard LSTM, which only processes the sequence forward in time, 
a BiLSTM maintains two parallel chains of hidden states: 
a forward chain $\overrightarrow{\mathbf{h}}_t$ that processes the sequence from $1 \rightarrow L$, 
and a backward chain $\overleftarrow{\mathbf{h}}_t$ that processes it in reverse from $L \rightarrow 1$. 
At each timestep, the two are concatenated as $\mathbf{h}_t = \big[\overrightarrow{\mathbf{h}}_t;\, \overleftarrow{\mathbf{h}}_t\big].$

This bidirectional setup is important because the meaning of an event often depends on its temporal context.  
For example, a brief period of missing data could be interpreted as a blink if it is followed immediately by normal gaze behavior, 
but the same gap might indicate disengagement if it occurs before and after long stretches of missingness.  
By looking both backward and forward in time, the BiLSTM can capture such context more effectively than a unidirectional model.  

\paragraph{Window-level representation and prediction.}
After processing a window of $L$ timesteps, the BiLSTM produces a sequence of hidden states 
$\{\mathbf{h}_t\}_{t=1}^L$, each encoding information about the input at time $t$ and its surrounding context.  
To obtain a fixed-length representation for the entire window, 
we concatenate the last forward hidden state $\overrightarrow{\mathbf{h}}_L$ (which summarizes information up to the end of the window) 
with the last backward hidden state $\overleftarrow{\mathbf{h}}_1$ (which summarizes information looking backward from the start): $\mathbf{z} = [\overrightarrow{\mathbf{h}}_L;\, \overleftarrow{\mathbf{h}}_1].$

This vector $\mathbf{z}$ serves as a compact summary of the whole sequence, 
capturing both past and future dependencies.  

The representation $\mathbf{z}$ is passed to a classification head, 
which applies a linear transformation followed by a softmax activation, outputting a probability distribution over the three categories for each personality trait. 

\paragraph{Training objective.}
Model training minimizes the categorical cross-entropy loss: $\mathcal{L} = -\sum_{c=1}^C \mathbf{1}[y=c] \,\log \hat{y}_c,$

which penalizes the divergence between the predicted distribution $\hat{y}$ and the true label $y$.  
Here, $\mathbf{1}[y=c]$ is an indicator function that equals $1$ when the ground-truth label is $c$ 
and $0$ otherwise. Intuitively, this loss encourages the model to assign high probability to the correct class 
while discouraging probability mass on incorrect categories.


\section{Experimental Evaluation}
\label{sec:experimental evluation}

\subsection{Classifier Training}
\label{sec:classifier-training}
We train a missing-data-aware \textit{bidirectional} LSTM classifier using a 12-dimensional input feature vector at each timestep: (1) horizontal gaze coordinate $g_x$, (2) its validity mask $m_{g_x}$, (3) temporal gap $\Delta t_{g_x}$, (4) vertical gaze coordinate $g_y$, (5) mask $m_{g_y}$, (6) gap $\Delta t_{g_y}$, (7) gaze velocity $v$, (8) mask $m_{v}$, (9) gap $\Delta t_{v}$, (10) pupil diameter $p$, (11) mask $m_{p}$, and (12) gap $\Delta t_{p}$. Sequences are segmented with a sliding window of length $L{=}100$ and overlap $=50$ (stride $=50$)~\cite{7134104}, which corresponds to a window duration of $1.67\,\text{s}$ with $0.83\,\text{s}$ stride at a $60\,\text{Hz}$ sampling rate.
The network uses hidden size $64$, $2$ layers, dropout $0.3$, and bidirectionality, followed by a two-layer classifier head; weights are initialized with orthogonal (LSTM) and Xavier (linear) schemes. Training uses cross-entropy loss with Adam optimization (learning rate $10^{-3}$, weight decay $10^{-5}$), gradient clipping, and a ReduceLROnPlateau scheduler (patience $=10$, factor $=0.5$), for up to $100$ epochs with early stopping after $15$ epochs without validation improvement. We evaluate performance using stratified 5-fold cross-validation (preserving the Low/Medium/High class distribution) and report mean accuracy and macro-averaged F1 scores with standard deviations. To prevent data leakage, we adopt a leakage-free data-splitting strategy in which data are split at the segment level: for each participant, continuous signals are first divided into non-overlapping contiguous segments, which are then assigned to one of the five folds. Sliding windows are generated only within segments after fold assignment, ensuring no overlap in raw data or windows between training and test folds.
 We also conduct participant-stratified cross-validation, assigning each participant entirely to one fold to test generalizability to unseen individuals, and report average macro F1 with standard deviations. Leave-one-subject-out (LOSO) was not feasible because each participant only provided data for one class. As a result, test folds contained a single class only, making accuracy and F1 computation infeasible.

\subsection{Results}
\label{sec:results}

\begin{table}[t]
\centering
\caption{Classification performance (mean $\pm$ standard deviation) across five folds for each Big Five trait using the full pipeline (time series + masks + temporal gaps). The last column reports participant-stratified cross-validation (by-participant) macro F1 averaged across folds. Values are percentages; the \% sign is omitted for readability.}
\vspace{-0.1in}
\small
\begin{tabular}{l c c c}
\toprule
Trait & Accuracy & Macro F1 & By-Participant F1 \\
\midrule
Openness          & $77.96 \pm 2.42$ & $77.52 \pm 2.63$ & $66.23 \pm \hspace{1ex}7.92$ \\
Conscientiousness & $74.86 \pm 2.34$ & $74.52 \pm 2.32$ & $65.99 \pm 25.12$ \\
Extraversion      & $78.17 \pm 1.49$ & $76.69 \pm 1.24$ & $70.89 \pm 13.92$ \\
Agreeableness     & $73.76 \pm 1.86$ & $73.09 \pm 2.05$ & $63.13 \pm 15.20$ \\
Neuroticism      & $79.19 \pm 2.70$ & $77.69 \pm 2.92$ & $74.33 \pm 23.85$ \\
\bottomrule
\end{tabular}
\vspace{-0.15in}
\label{tab:bigfive_results}
\end{table}

\begin{table*}[t]
  \centering
  \caption{Ablation study results across Big Five traits. ``Full'' uses time series + masks + temporal gaps features; ``TS+Temporal Gap'' uses time series + temporal gap features; ``TS Only'' uses raw time series features; and ``Statistical'' uses handcrafted statistical features with a Random Forest. Results are reported as mean accuracy and macro F1 (\%) with standard deviation.}
  \small
  \begin{tabular}{l c c c c c}
     & Openness & Conscientiousness & Extraversion & Agreeableness & Neuroticism \\
    \toprule 
    \multicolumn{6}{c}{Accuracy (\%)}\\
    \midrule
    Full & \cellcolor[HTML]{caebc0}\textbf{77.96 $\pm$ 2.42}
         & \cellcolor[HTML]{caebc0}\textbf{74.86 $\pm$ 2.34}
         & \cellcolor[HTML]{caebc0}\textbf{78.17 $\pm$ 1.49}
         & \cellcolor[HTML]{caebc0}\textbf{73.76 $\pm$ 1.86}
         & \cellcolor[HTML]{caebc0}\textbf{79.19 $\pm$ 2.70} \\

    TS+Temporal Gap
         & 69.96 $\pm$ 1.50
         & 70.43 $\pm$ 1.48
         & 73.49 $\pm$ 2.65
         & 67.15 $\pm$ 1.70
         & 71.80 $\pm$ 0.99 \\

    TS Only
         & 69.23 $\pm$ 2.48
         & 66.86 $\pm$ 0.74
         & 72.67 $\pm$ 2.32
         & 65.75 $\pm$ 1.36
         & 73.44 $\pm$ 0.44 \\

        Statistical
         & 67.59 $\pm$ 1.84
         & 67.15 $\pm$ 2.18
         & 70.97 $\pm$ 3.51
         & 65.56 $\pm$ 1.67
         & 73.73 $\pm$ 2.08 \\

    \midrule
    \multicolumn{6}{c}{Macro F1 Score (\%)}\\
    \midrule
    Full & \cellcolor[HTML]{caebc0}\textbf{77.52 $\pm$ 2.63}
         & \cellcolor[HTML]{caebc0}\textbf{74.52 $\pm$ 2.32}
         & \cellcolor[HTML]{caebc0}\textbf{76.69 $\pm$ 1.24}
         & \cellcolor[HTML]{caebc0}\textbf{73.09 $\pm$ 2.05}
         & \cellcolor[HTML]{caebc0}\textbf{77.69 $\pm$ 2.92} \\

    TS+Temporal Gap
         & 69.87 $\pm$ 1.54
         & 70.38 $\pm$ 1.52
         & 70.29 $\pm$ 3.12
         & 66.21 $\pm$ 1.68
         & 69.60 $\pm$ 2.58 \\

    TS Only
         & 68.73 $\pm$ 2.45
         & 66.51 $\pm$ 0.71
         & 70.64 $\pm$ 2.12
         & 65.56 $\pm$ 1.46
         & 71.56 $\pm$ 0.37 \\

    Statistical
         & 66.40 $\pm$ 1.80
         & 66.99 $\pm$ 2.27
         & 67.85 $\pm$ 3.30
         & 64.62 $\pm$ 1.57
         & 70.31 $\pm$ 2.90 \\

    \bottomrule
\end{tabular}

  \label{tab:ablation_results}
\end{table*}

\begin{figure*}[htbp]
    \centering
    \begin{subfigure}{0.19\textwidth}
        \centering
        \includegraphics[width=\linewidth]{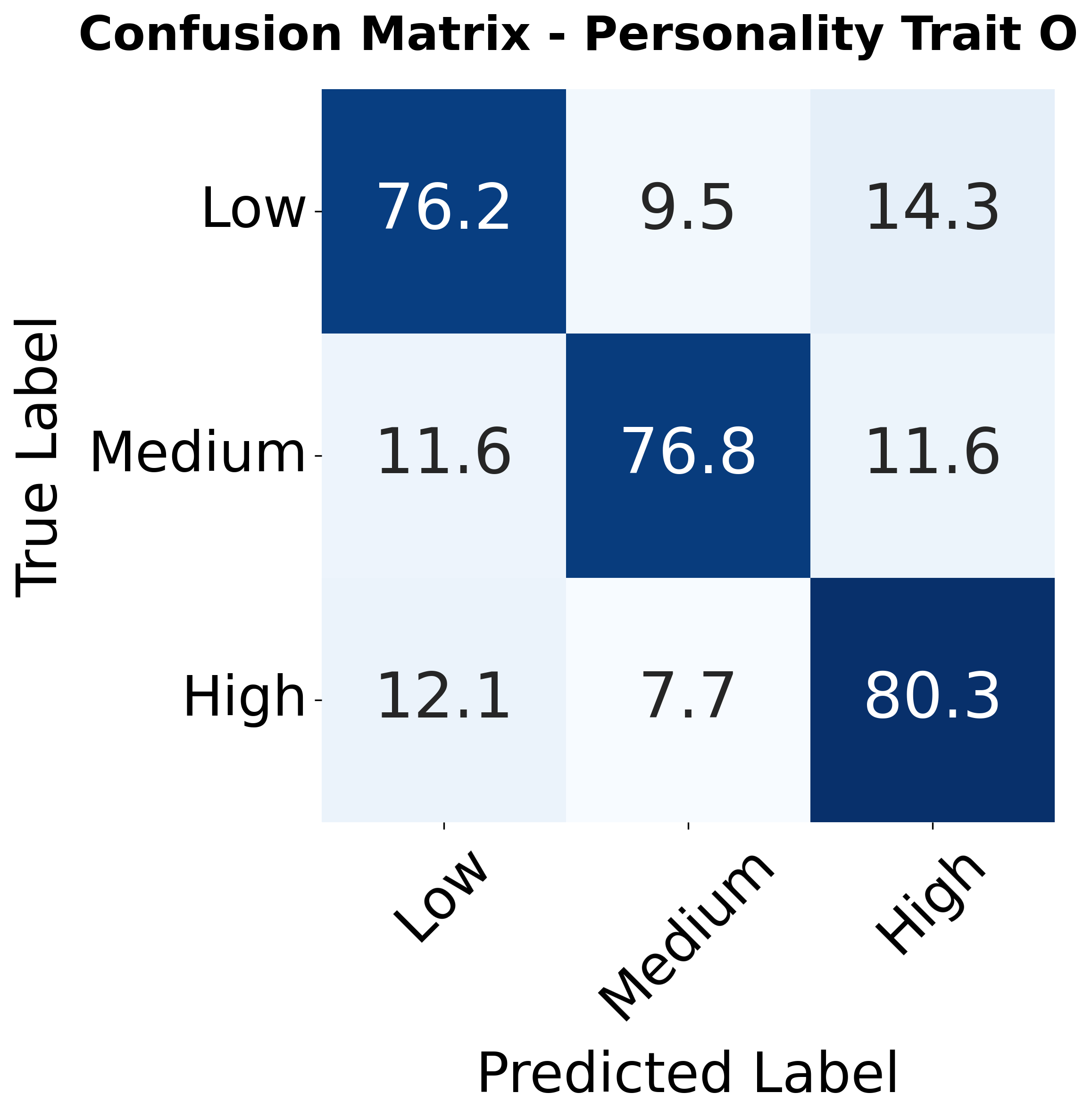}
        \caption{O}
    \end{subfigure}
    \begin{subfigure}{0.19\textwidth}
        \centering
        \includegraphics[width=\linewidth]{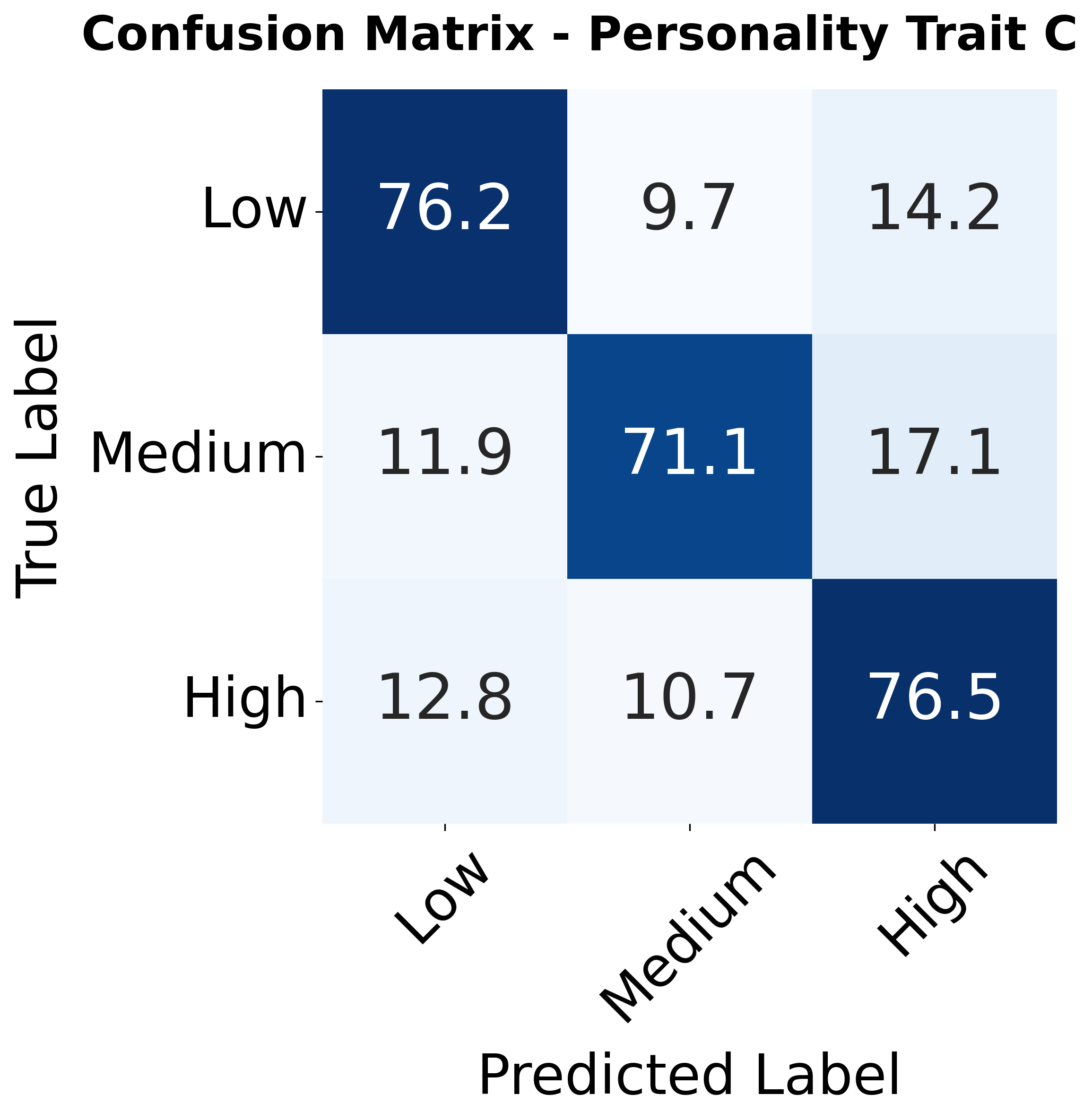}
        \caption{C}
    \end{subfigure}
    \begin{subfigure}{0.19\textwidth}
        \centering
        \includegraphics[width=\linewidth]{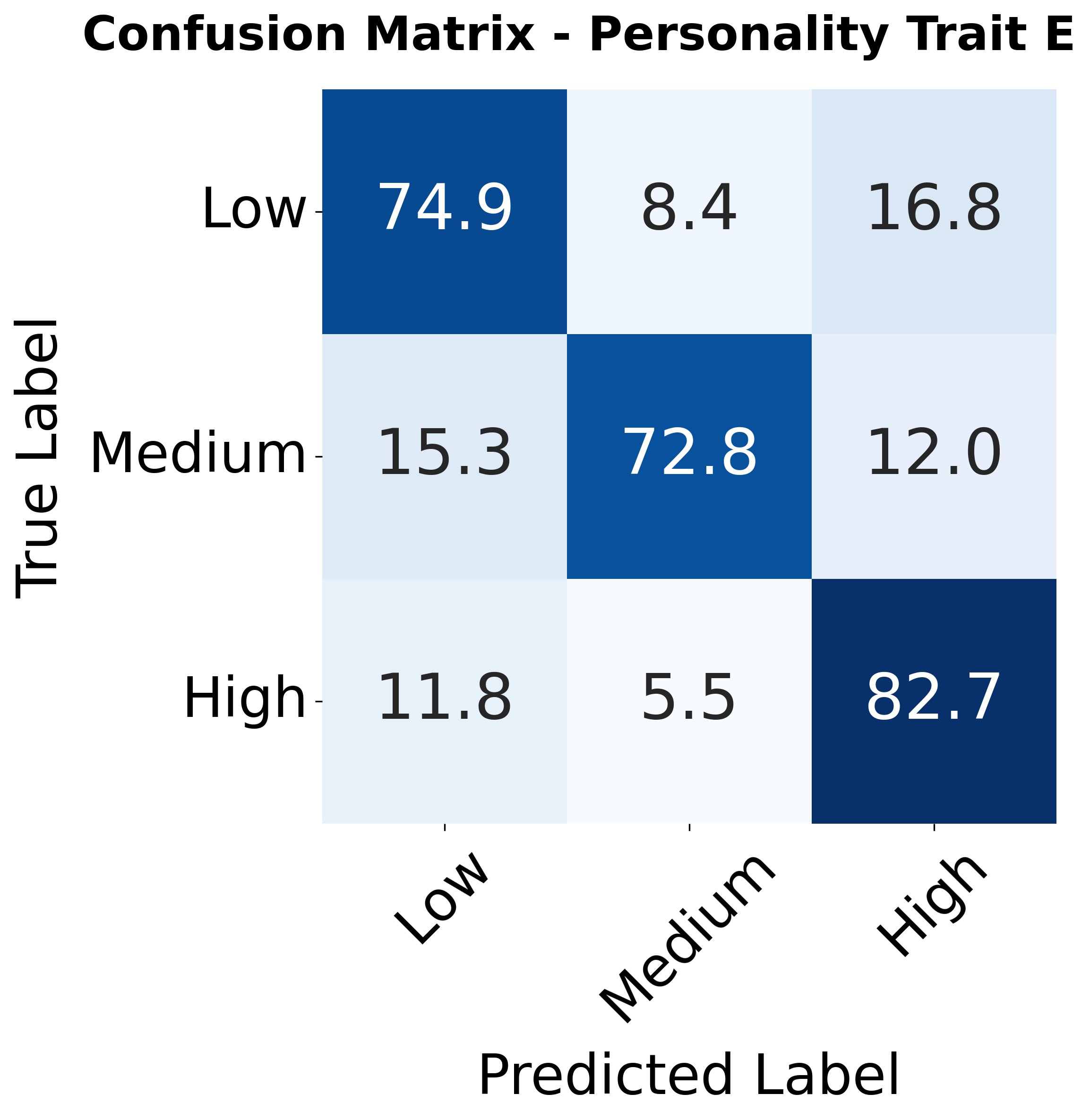}
        \caption{E}
    \end{subfigure}
    \begin{subfigure}{0.19\textwidth}
        \centering
        \includegraphics[width=\linewidth]{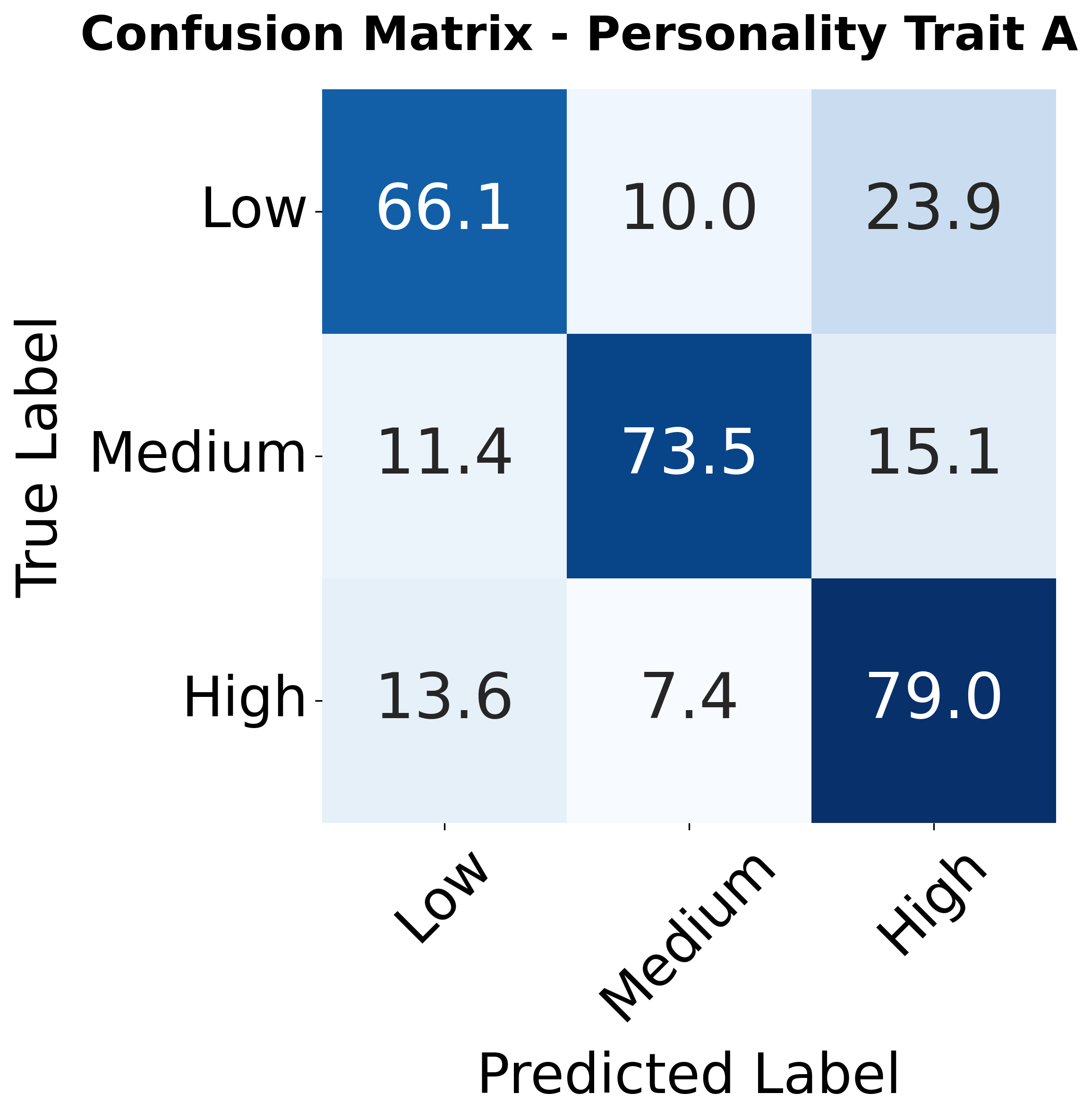}
        \caption{A}
    \end{subfigure}
    \begin{subfigure}{0.19\textwidth}
        \centering
        \includegraphics[width=\linewidth]{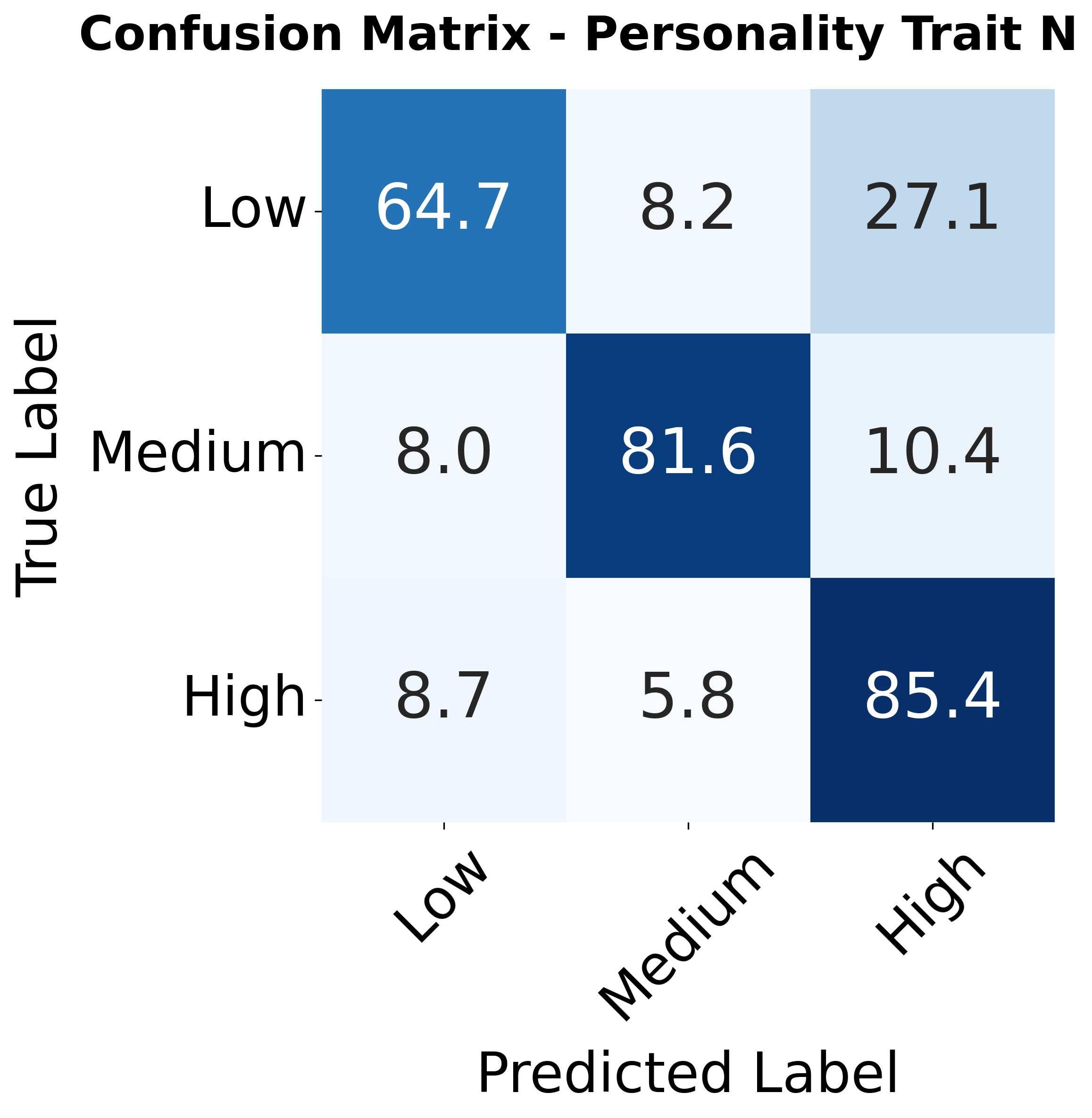}
        \caption{N}
    \end{subfigure}
    \vspace{-0.1in}
    \caption{Confusion matrices for each personality trait.}
    \vspace{-0.15in}
    \label{fig:five-in-one-row}
\end{figure*}

Table~\ref{tab:bigfive_results} summarizes the classification performance for each of the Big Five traits. Overall, the proposed missing-data-aware BiLSTM achieves stable and moderately strong performance across all traits, with mean accuracies ranging from $73.09\%$ (Agreeableness) to $77.69\%$ (Neuroticism). Neuroticism ($77.69\%$) and Openness ($77.52\%$) exhibited relatively higher predictive performance. Extraversion ($76.69\%$) and Conscientiousness ($74.52\%$) also achieved competitive results. Agreeableness remains the most challenging trait, with macro F1 score at $73.09\%$. Figure~\ref{fig:five-in-one-row} depicts the confusion matrices for each of the Big Five traits. 

In addition to fold-level cross-validation, we evaluated performance using participant-stratified cross-validation to test generalizability to unseen individuals. While performance naturally decreased in this stricter setting (e.g., Macro F1 score of $63.1\%$ for Agreeableness and $66.2\%$ for Openness), the model still achieved reasonable performance, with Extraversion ($70.9\%$) and Neuroticism ($74.3\%$) showing the strongest generalization.

\subsection{Ablation Study}
\label{sec:ablation}

To examine whether incorporating missingness information affects the model performance, we conducted a set of ablation experiments where different subsets of features and modeling strategies were retained.
Here, binary masks denote missingness indicators (1 = valid, 0 = missing).

\begin{enumerate}
    \item \textbf{Full Pipeline.} The proposed missing-data-aware BiLSTM trained with the complete representation: time-series features (horizontal/vertical gaze, pupil diameter, gaze velocity), their binary validity masks, and temporal gap encodings.
    \item \textbf{Time Series + Temporal Gap.} The BiLSTM is trained using the four time-series features together with temporal gap ($\Delta t$) encodings, but without missingness indicators. In this setting, the model retains information about the elapsed time since the last observation, but does not explicitly encode whether a value is observed or missing.
    \item \textbf{Time Series Only.} The BiLSTM trained solely on the four raw time-series features, without masks or temporal gaps. This serves as a baseline for sequential modeling without any explicit missingness indicators.
    \item \textbf{Non-sequential Baseline. } A non-sequential model trained on handcrafted features. For each of the four base signals (gaze $x$, gaze $y$, pupil, gaze velocity), we compute five descriptive statistics (minimum, maximum, mean, standard deviation, median), yielding a 20-dimensional feature vector. A random forest classifier is then trained on this representation, providing a feature-engineered baseline without temporal modeling.
\end{enumerate}

Table~\ref{tab:ablation_results} summarizes the results of the ablation study across the Big Five traits. The full pipeline, which combines time-series features, explicit missingness masks, and temporal gap encodings, consistently achieves the best performance across traits, with accuracies ranging from approximately $74\%$ (Agreeableness) to $79\%$ (Neuroticism) and macro F1 scores between $73\%$ and $78\%$. Removing missingness masks while retaining temporal gap encodings (\emph{TS+Temporal Gap}) leads to consistent performance degradation across all traits, suggesting that explicit missingness indicators provide complementary information beyond temporal gap encodings alone. Using only raw time-series features (\emph{TS Only}) further reduces performance, indicating that modeling both irregular sampling and missing data is important for effective temporal representation learning. The statistical baseline, which relies on handcrafted features and a Random Forest classifier, performs worst overall, highlighting the benefit of sequential models augmented with explicit temporal and missingness representations. Overall, the ablation results demonstrate that each component of the proposed pipeline contributes meaningfully to robust personality trait prediction.

\vspace{-0.1in}
\section{Discussion}
\label{sec:discussion}

The results in Table~\ref{tab:bigfive_results} show that it is indeed possible to classify personality traits from temporal behavioral features, and that this approach works best when the full modeling pipeline is used -- that is, when raw time-series signals are combined with missingness and temporal gap information. The consistently strong scores suggest that eye-tracking data contains rich signals about stable psychological traits. At the same time, the differences in performance across the Big Five traits reveal how personality is expressed unevenly in gaze behavior. These patterns provide not only theoretical insights into the behavioral expression of personality but also practical guidance for how to build computational models.

Among the five traits, \textit{Neuroticism (N)} and \textit{Openness (O)} showed relatively stronger and more consistent performance, in line with prior work linking these traits to distinctive gaze variability and exploratory viewing behaviors~\cite{rauthmann2012eyes,perlman2009individual,agnoli2015eye,risko2012curious}. \textit{Extraversion (E)} and \textit{Conscientiousness (C)} achieved moderate performance, suggesting that attentional cues are informative but less distinctive in solitary tasks~\cite{le2024way,hoppe2018eye,tsigeman2024ai}. In contrast, \textit{Agreeableness (A)} remained the most difficult trait to predict, consistent with evidence that it is primarily expressed in social interaction contexts~\cite{rauthmann2012eyes,wu2014influence}. Overall, traits tied to internal regulation (attention, emotion) were easier to detect from gaze than socially oriented traits, which may require richer contexts or additional modalities.

When examining generalization, the by-participant cross-validation revealed a different picture. Scores were lower and more variable across individuals, especially for Agreeableness and Conscientiousness. This highlights a major challenge: while group-level models capture broad tendencies, predicting personality consistently across different individuals is more complex. One reason is the heterogeneity of cognitive strategies: individuals may differ in how they allocate attention, process information, or regulate effort during a task. For example, some people naturally adopt systematic scanning routines, while others rely on more opportunistic or intuitive search behaviors~\cite{hoppe2018eye, berkovsky2019detecting}, leading to divergent gaze dynamics even if they share the same personality trait~\cite{macfarlane2017visual}. Another possible explanation is the limited sample size: with only 25 participants, the distribution of Big Five personality traits may differ substantially from that observed in much larger populations~\cite{srivastava2003development}. This discrepancy can influence performance at the individual-participant level and limit the model’s ability to generalize. Increasing the number and diversity of participants would likely lead to a more representative trait distribution and improve generalization performance.

Behavioral habits and situational factors further complicate generalization. Gaze patterns are shaped not only by personality but also by momentary states such as fatigue, stress, or task engagement~\cite{zargari2018eye, fleeson2009implications}. A conscientious individual (Conscientiousness) might display high attentional stability in one session but exhibit more variability under distraction or cognitive load. Similarly, Agreeableness -- already subtle in solitary tasks -- may only surface in socially interactive contexts, making it less consistently observable across participants~\cite{wu2014influence}.

These considerations suggest that variability is not simply ``noise'' but reflects meaningful person–context interactions. Capturing these nuances may require models that account for both stable dispositions and dynamic behavioral states. Approaches such as domain adaptation, personalized modeling, or hierarchical frameworks could help bridge the gap, enabling models to distinguish between trait-driven regularities and state-driven fluctuations. More broadly, these findings underscore the importance of integrating cognitive and behavioral theories into computational modeling: personality is not expressed uniformly, but filtered through the individual’s strategies, habits, and situational context~\cite{fleeson2009implications}.

The ablation results in~\autoref{tab:ablation_results} emphasize why temporal context matters. The complete pipeline (time series + masks + temporal gaps) consistently outperformed both raw time series alone and handcrafted statistical features. Notably, the inclusion of masks and temporal gap information added substantial value. This suggests that personality-relevant information is not just in the sequence of gaze coordinates but also in irregularities -- when people look away, how long gaps last, and how missing data is structured. This aligns with recent research in time-series modeling that treats missingness as informative rather than as
noise~\cite{lipton2016modeling, che2018recurrent}. Ignoring such irregularities risks discarding psychologically meaningful signals.

From a cognitive perspective, irregularities reflect the underlying processes of attention and information processing. Long gaps can signal lapses in sustained attention or shifts in cognitive focus, often linked to high Neuroticism or lower engagement with the task~\cite{doherty2005gaze}. \citet{doherty2005gaze} noted that people tend to look away when under high cognitive load. By contrast, more frequent gaze shifts and less tightly constrained fixation patterns may reflect exploratory viewing strategies, which are characteristic of individuals high in Openness, who tend to engage flexibly with stimuli and seek novel information. Thus, the timing of gaze interruptions offers indirect but meaningful cues about attention and control.

Behaviorally, missingness patterns capture differences in task approach. Some individuals show bursts of exploratory scanning interspersed with pauses -- consistent with curiosity and Openness -- while others show repetitive fixations and fewer breaks, reflecting more rigid or habitual engagement~\cite{agnoli2015eye}. This means irregular timing is not just a random error but a behavioral marker of personality-linked cognitive styles.

The methodological implications are equally important. Traditional feature engineering often assumes that missing data should be interpolated or discarded, but our findings indicate that the very presence and distribution of missingness can serve as predictive features. This perspective aligns with a growing body of behavioral informatics research, which argues that ``what is absent'' in behavioral traces can be as informative as ``what is present''~\cite{che2018recurrent}. For personality modeling, this means that gaze dynamics should be analyzed not only for their visible sequences but also for their absences, interruptions, and irregular rhythms.

In summary, our findings show that eye-tracking holds promise for inferring personality, albeit with important boundaries. Traits tied to attention and emotion, such as Openness and Neuroticism, are especially well captured, while socially expressed trait like Agreeableness may require richer data sources such as language, physiology, or social interaction~\cite{taib2020personality,tsigeman2024ai}. Task design also plays a critical role: social or collaborative contexts may amplify expressions of traits that remain muted in solitary tasks.

In general, the temporal structure of eye movements, including where gaps occur, provides meaningful cues about personality. However, not all traits are equally predictable, and performance varies across individuals. Recognizing these nuances is crucial both for advancing personality science and for building adaptive systems that personalize interactions. Combining eye-tracking with multimodal data and more diverse task designs could lead to models that are both more comprehensive and more generalizable.

\section{Conclusion}
\label{sec: conclusion}

This study demonstrates that eye-tracking data, especially when analyzed as time-based sequences that include gaps and missing points, can provide useful insights into personality. We found that traits related to attention and emotion regulation, such as Neuroticism and Openness, are captured well in gaze patterns. In contrast, socially driven traits like Agreeableness is harder to detect in tasks performed alone. This highlights both the promise and the limits of using eye movements alone to infer personality.

Beyond prediction accuracy, our work makes two broader contributions. First, we demonstrate that irregularities in gaze, moments when people look away or pause, are not just noise but carry meaningful signals about attention control, disengagement, and exploration. Second, we highlight that personality is shaped by both stable traits and changing contexts, which means models must account for individual differences while also adapting to situational influences.

\myparagraph{Limitations} Our study focused exclusively on museum settings using picture-based search systems. Additionally, different eye-tracking devices may yield varying results due to differences in processing capabilities and other influencing factors, such as sampling frequency, camera resolution, and calibration quality. 

\myparagraph{Future Work} Future research could extend this work to a wider range of interactive settings. Additionally, combining eye-tracking data with other modalities such as electrodermal activity (EDA), electroencephalography (EEG), or electrocardiography (ECG) could enrich multimodal models and offer deeper insights into how personality traits influence information-seeking behavior.

\newpage
\begin{acks}
This research is supported by National Institute of Informatics (NII) Internship Program, the \grantsponsor{JSPS}{Japan Society for the Promotion of Science}{https://www.jsps.go.jp/english/} (JSPS, \grantnum{JSPS}{23K28375}), and the \grantsponsor{ARC}{Australian Research Council (ARC)}{https://www.arc.gov.au/} Centre of Excellence for Automated Decision-Making and Society (ADM+S, \grantnum{ARC}{CE200100005}).
\end{acks}

\balance
\bibliographystyle{ACM-Reference-Format}
\bibliography{00-refs}

\end{document}